\documentclass{vgtc}                          

\ifpdf
  \pdfoutput=1\relax                   
  \usepackage{graphicx}                
  \DeclareGraphicsExtensions{.pdf,.png,.jpg,.jpeg} 
\else
  \usepackage{graphicx}                
  \DeclareGraphicsExtensions{.eps}     
\fi%

\graphicspath{{figures/}{figures/}{images/}{./}} 

\usepackage{microtype}                 
\PassOptionsToPackage{warn}{textcomp}  
\usepackage{textcomp}                  
\usepackage{mathptmx}                  
\usepackage{times}                     
\usepackage{cite}                      
\usepackage{tabu}                      
\usepackage{booktabs}                  
\usepackage{float}

\usepackage{enumitem}
\setlist[itemize]{align=parleft,left=0pt..1em}

\onlineid{0}

\vgtccategory{Research}

\vgtcinsertpkg

\title{Smell of Fire Increases Behavioural Realism in Virtual Reality: A Case Study on a Recreated MGM Grand Hotel Fire}

\author{Humayun Khan\thanks{e-mail: humayun.khan@canterbury.ac.nz} %
\and Daniel Nilsson }
\affiliation{\scriptsize VR Evacuation Lab, CNRE, University of Canterbury}

\abstract{

Virtual reality allows creating highly immersive visual and auditory experiences, making users feel physically present in the environment. This makes it an ideal platform to simulate dangerous scenarios, including fire evacuation, and study human behaviour without exposing users to harmful elements. However, human perception of the surroundings is based on the integration of multiple sensory cues (visual, auditory, tactile, or/and olfactory) present in the environment. When some of the sensory stimuli are missing in the virtual experience, it can break the illusion of being there in the environment and could lead to actions that deviate from normal behaviour. In this work, we added an olfactory cue in a well-documented historic hotel fire scenario that was recreated in VR, and examined the effects of the olfactory cue on human behaviour. We conducted a between subject study on 40 naive participants. Our results show that the addition of the olfactory cue could increase behavioural realism. We found that 80\% of the studied actions for the VR with olfactory cue condition matched the ones performed by the survivors. In comparison, only 40\% of the participants' actions for VR only condition were similar to the survivors.

} 

\CCScatlist{
  \CCScatTwelve{Olfactory Augmentation}{Human Behaviour}{Virtual Reality}{-}{}
  \CCScatTwelve{Fire Evacuation}{Emergency Response}{}{}
}
     
\begin{document}


\firstsection{Introduction}
\maketitle
Understanding human behaviour during a fire is essential to design buildings and procedures to safely and efficiently evacuate people. To examine the behaviour of people during a fire evacuation, multiple approaches (observational and simulated) have been tried in the past, including hypothetical experiments, case studies, field studies, drills (announced and unannounced), laboratory physical and virtual reality (VR) experiments \cite{arias2019forensic, lovreglio2020augmented, kinateder2014virtual}. Of these methods, VR has emerged as a promising tool to study human behaviour in fire. It provides a medium to immerse participants in virtual scenarios and environments that would be hazardous and impossible to replicate in the real-world. By simulating fire incidents in VR, researchers can create realistic and dynamic scenarios that closely resemble actual emergency situations.

The immersion in the virtual scenarios can evoke physical, psychological, and physiological responses from participants as they would experience in a real fire incident. This allows researchers to examine human behaviour, decision-making, and evacuation strategies in a controlled and safe environment. Participants can go through virtual buildings, encountering different hazards, and making critical decisions. By studying these behaviours in VR simulations, researchers can gather valuable insights on the factors affecting evacuation, such as crowd dynamics, decision-making under stress, evacuation times, and compliance with emergency protocols. These outcomes can then be used to improve architectural designs, enhance training programs, and develop more effective emergency procedures for occupants and emergency responders. Furthermore, the knowledge gained through VR experiments can also inform the development and improvement of fire safety regulations and guidelines. By understanding people's behaviour in virtual fire scenarios, building designers and decision-makers can identify weaknesses in building designs or emergency protocols and make necessary adjustments to enhance safety measures.

Even though VR provides a physically immersive medium, it is not a substitute of real-world fire incidents. However, it provides a controlled environment to study human behaviour during emergencies, offering valuable information that is difficult to obtain otherwise. Integrating findings from diverse research methods, including VR experiments, with real-world data broadens our understanding of human behaviour during fire evacuations. This in turn can improve our current fire safety practices.

\vspace{1em}
\noindent This paper makes the following contributions:
 \begin{itemize}
     \item Presents a user study investigating the effect of smell stimuli on human behaviour and perception in virtual reality for an evacuation scenario.
     \item Introduces a novel smell delivery mechanism that passively displays the olfactory cue.
 \end{itemize}
\section{Related Work}
This section summarises past research on human behaviour in fire, and how VR has been used as a tool of investigation. Additionally, it outlines previous work on olfactory technologies for VR - their smell generation and delivery mechanism, and limitations.

\subsection{Human Behaviour in Real Fire Evacuation Scenarios}
Human behaviour in fire (HBiF) has been studied extensively in the past research encompassing survivors' data \cite{bryan1983review, bryan1983examination}, real-world incident reports \cite{best1982investigation, kuligowski2016human},  behavioural models \cite {kuligowski2008modeling}, and laboratory experiments \cite{arias2019forensic, shaw2019heat, lovreglio2020augmented}. These studies have improved our understanding of the field and have been summarised in existing reviews \cite{bryan1999human, kinateder2015risk, cheng2019human}. This paper is not intended to enhance our overall understanding of the HBiF, but to further advance the use of VR as a tool for research. It particularly focuses on the use of olfactory cue and its effects on human behaviour. Fire cues and pre-evacuation are the main concepts from HBiF that concern this paper \cite{kuligowski2008modeling, kinateder2015risk}. People with higher risk perception exhibit behaviours that are aligned with their perception of the environment and respond more quickly to evacuation scenarios than individuals who ignore emergency scenarios \cite{kinateder2015risk}. Fire cues play an important role in initiating the evacuation or fire suppression process. More cues and proximity to the fire will lead to a faster response \cite {kuligowski2008modeling}. During pre-evacuation, people often engage in non-evacuation behaviours, such as gathering belongings, changing clothes, making phone calls, shutting down computers, or seeking permission to leave; these actions lead to delay and increase their risk of getting hurt.

To validate the human behaviour shown during the VR experiment is closer to the real-life evacuee behaviour, a well-documented real-world fire incident is needed. The MGM Grand hotel fire that occurred in November, 1980 meets the criteria \cite{best1982investigation}. The fire started on the ground floor in a restaurant “the Deli” due to an electrical fault. As there were few fire safety measures in place, the fire spread rapidly throughout the hotel building. Due to the flammable building material and hotel’s ventilation system, the fire spread horizontally and vertically, trapping guests and employees in their rooms. There was thick smoke in the hallways and staircases that made it difficult to see. There was also lack of emergency lighting and signage that complicated way-finding. The alarm system did not work in the main hotel tower that prevented the people from getting timely warning. This lead to people being trapped in their rooms specially on the higher floors. These trapped individuals perform actions to manage the smoke and survive until they were rescued by the firefighters. From an estimated 5000 occupants, 85 people died and most of the fatalities were due to the smoke inhalation. After the fire a thorough investigation was conducted by Bryan \cite{bryan1982human} to analyse the behaviours of survivors during the fire. The data was collected by the National Fire Protection Association (NFPA) through a mailed survey to the survivors. By comparing the observed behaviours of the participants during the VR experiments to the NFPA survey’s survivor data, we will be able to validate which of the participants actions match the real survivors actions. Our objective is to further advance the use of VR as a research tool to analyse human behaviour in fire. 

\subsection{Studies using VR to Investigate Fire Evacuation}
Numerous studies have used VR to simulate fire emergencies and examine different aspects of evacuation behaviours, including way-finding \cite {arias2019virtual_wayfinding, shaw2019heat}, egress behaviours \cite{arias2021virtual}, exit signage \cite{olander2017dissuasive}, and elevators waiting time in high-rise buildings \cite{mossberg2021evacuation}. VR-based fire evacuation simulations provide a safe and ethical medium to design and conduct experiments that closely resembles the real-world environment. In VR, researchers can expose participants to dangerous hypothetical or real scenarios without risking their safety. This allows researchers to identify different evacuation strategies, evaluate the effectiveness of safety measures, find pre-evacuation response time for different demographics, and determine participants’ movement speed and routes taken during the evacuation. These insights into crowd dynamics help identify evacuation bottlenecks and formulate strategies for enhancing evacuation speed.

While VR headsets create a believable visual and auditory experience, they lack sensory modalities, such as smell, touch, heat, and taste. This absence  might have caused participants in the studies \cite{arias2019forensic, chalmers2021realistic, shaw2019heat} to exhibit behaviours inconsistent with reality. Researchers have argued that including these sensory inputs (especially tactile and olfactory cues) would increase behavioural realism shown by the participants as human perception of the real-world is based on the integration of all the sensory feedback in the environment \cite{cheok2011multi}. This aligns with HBiF research, indicating a higher number of fire cues leads to higher risk perception that causes evacuees to take action \cite{kinateder2015risk}. However, there is limited research conducted on adding multisensory components in VR and studying evacuation behaviour, mainly due to the lack of established hardware to deliver these sensory modalities. Past research on including heat and smell in evacuation scenarios concluded that the stimuli lacked realism and discrepancy in synchronisation with the visuals \cite{shaw2019heat}. This  presents an opportunity to add these stimuli with sufficient fidelity to stimulate the senses, and also remove the stimuli when it is no longer needed.


\subsection{Displaying Smell in the Virtual Reality} \label{smellDisplays_LitReview}
Many olfactory devices have been made in the past to deliver smell in VR \cite{niedenthal2023graspable, bahremand2022smell, de2022olfactory, brooks2020trigeminal, maggioni2018smell, ranasinghe2018season, Liu2023, yanagida2004projection}. These devices involve mainly two mechanisms: smell generation and smell delivery. During smell generation the odour vaporises from its stock form (e.g. essential oil) to an air mixture, and delivered to the human olfactory organ(nose) with the smell delivery mechanism. The smell generation can be actively done using heating or atomisation, or can be passive through natural vaporisation. The smell delivery mechanism that have been used in the past were based on a fan blowing the smell air mixture to the user wearing VR headset \cite{ranasinghe2018season, ariyakul2011improvement}, or air canon shooting smell vortex to the user nose \cite{hu2021abio, yanagida2004projection}, or tubes bringing the smell from source to the user \cite{bahremand2022smell, brooks2020trigeminal, ranasinghe2018season, maggioni2018smell}, or smell generator placed just below the user’s nose \cite{brooks2020trigeminal, Wang2020_OnFace, Liu2023}.  Each device has their own limitation, but the common themes emerging from these smell delivery mechanisms are long lingering times (smell present for some time after the first release), inaccurate delivery trajectories, delivery time, noise of the device, and cumbersome setup that has to be worn by the user that limits their mobility in the space.

\section{User Study}
This work studies how presence of an olfactory cue (fire smell) affects an individual behaviour and perception during an emergency scenario. It answers the question whether olfactory cue leads to a more realistic behaviour exhibited by participants. It also introduces a novel smell delivery mechanism. For the reference data, we selected a well-documented real-world fire incident (MGM Grand hotel fire) with validated actions identified by the survivors and collected by NFPA \cite{bryan1983review}. We conducted our experiment exposing only one participant at a time to a hotel room evacuation scenario and observed their behaviour in response to the incident. To know the realism of the behaviour, we compared participant's actions during the experiment to the survivors and saw how closely they matched. We also had a between-subject condition in which participants were only exposed to VR without the smell. To limit the effect of social influence, there were no other building occupants in sight. 

\subsection{Our Smell Delivery System}  
\begin{figure}[H]
 \centering 
\includegraphics[width=0.8\columnwidth]{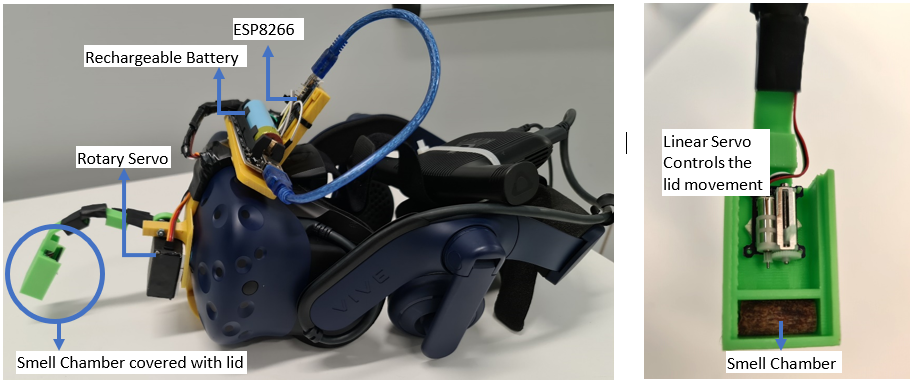}
 \caption{ Passive olfactory display that was used to deliver the smell. }
 \label{POD_Display}
\end{figure}

To overcome the issues mentioned in section \ref{smellDisplays_LitReview}, specially lingering smell, delivery time, and portability, we created the passive olfactory display (POD) that introduces the smell in VR without actively emitting smell molecules in the air. Our device allows more precise spatio-temporal control of smell delivery in the VR, and it is also wireless and portable. Figure \ref{POD_Display} shows our device. It has a smell chamber covered by a sliding lid that is attached to a linear actuator to slide open and close the chamber. The chamber is at the end of a 3D-printed arm that is actuated with a rotary servo motor. The physical dimensions of the smell chamber are L:12.1 x W:17.9 x H:13.2 mm, the smell container are L:46.6 x W:24.0 x H:18.3 mm, and the arm are L:135.2 x W:60.5 x H:88.7 mm. Both the linear actuator and rotary servo motor are controlled with PWM signals from a microcontroller (ESP8266) that is wirelessly linked to the PC running the VR simulation. The smell inside the chamber is on a cotton wick with birch tar essential oil. When the smell needs to be delivered in VR, signals are sent from the PC to the microcontroller that moves the arm closer to the user’s nose and opens the lid, the ON state. Similarly, when the smell has to be removed, the PC signals the microcontroller that closes the lid and moves the arm away from the user’s nose, OFF state. There were only two operational states, ON (smell activation) state and OFF (smell deactivation state). Also, the arm movement happens in real-time with the going down motion taking 0.7 seconds and the lid opening happens in 0.45 seconds. This allows quick introduction and removal of the smell.

\subsection{Study Design}
\begin{figure}[htb]
 \centering 
\includegraphics[width=\columnwidth]{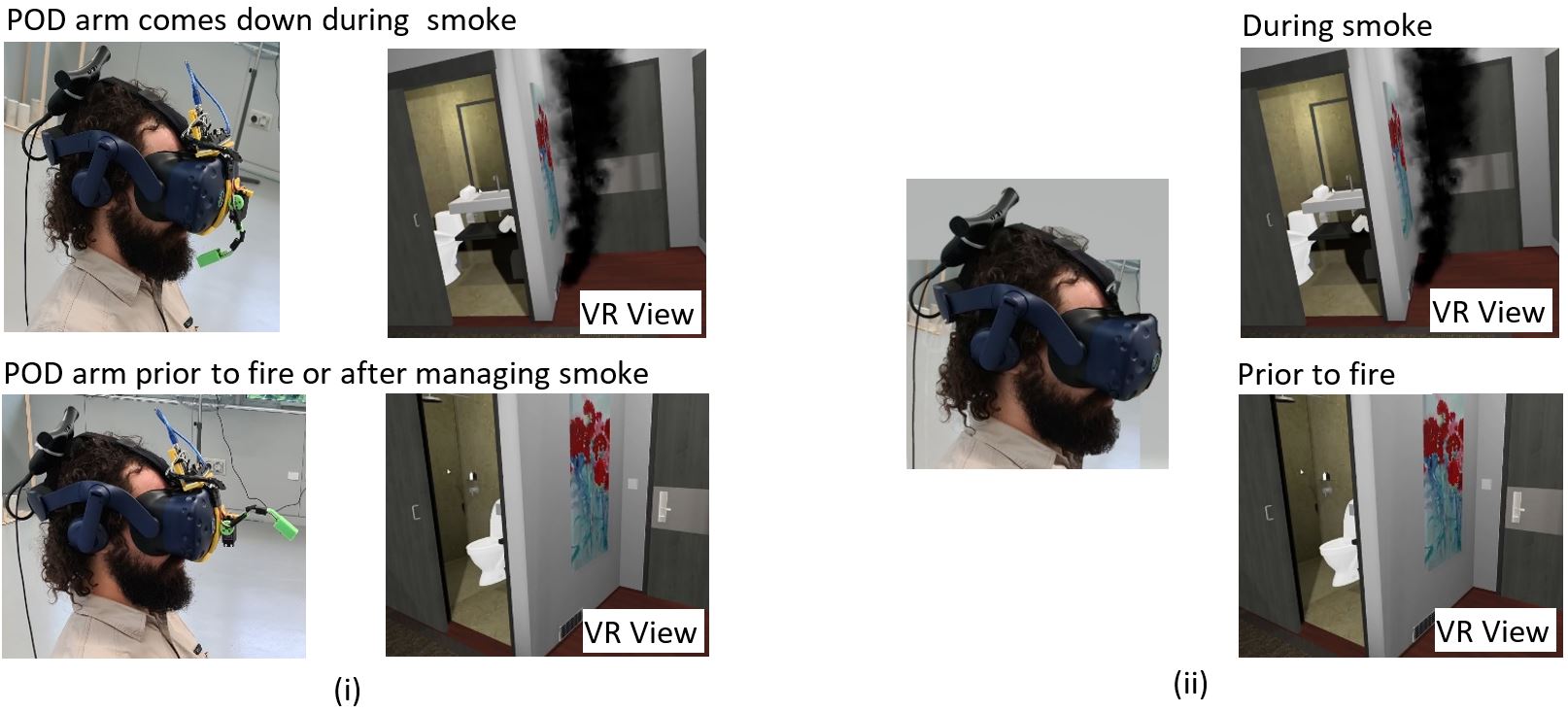}
 \caption{Hotel room fire scenario in VR, user study conditions (i) with smell, and (ii) without smell.}
 \label{UserStudyDesign_Photo}
\end{figure}

We conducted a between-subject experiment with two conditions: i) hotel room fire scenario in VR with smell (VR with smell condition), and ii) the same scenario in VR without smell (VR only condition). The factorial design of our experiment (2x1) is illustrated in Table \ref{FactorianDesign}. Each participant only experienced one condition. Prior to the main experiment, all participants went through the training phase which lasted for about 5 minutes. In the training phase, participants were familiarised with how to interact with objects in a training VR environment using HTC Vive hand controllers and the available physical space they can move in during the experiment. For the main experiment, there were 40 participants, 20 per condition. The hotel room VR environment and the possible interactions were the same for both conditions. Both conditions are shown in Figure \ref{UserStudyDesign_Photo}. 

\begin{table}[htb]
 \centering 
 \caption{Factorial design of the user study}
 \includegraphics[width=0.8\columnwidth]{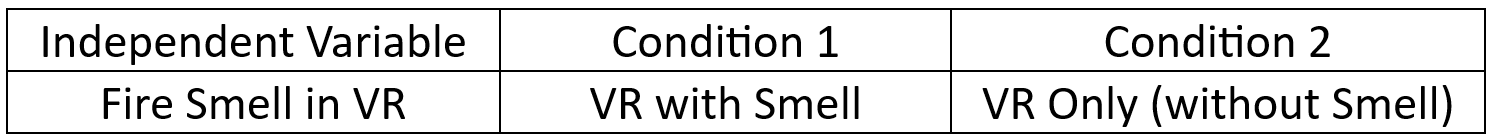}
 \label{FactorianDesign}
\end{table}

\vspace{-1em}
\subsection{Research Objectives}
Following are our research objectives for this work:

\begin{itemize}
    \item \textbf{O1}: To examine whether adding fire smell in VR  results in more realistic behaviour shown by the participants in the simulated hotel fire scenario. Realistic behaviour for our study is the actions performed by the participants similar to the actions of real-life survivors shown in the NFPA data. This is analysed by individually comparing the actions performed data of the two conditions (VR with smell, VR only) and the real survivors' NFPA data.
    \item \textbf{O2}: To investigate if there is any change in participants' behaviour due to the addition of fire smell. This is done by comparing the actions performed data between the VR with smell and VR only condition.
    \item \textbf{O3}: To determine if the different demographics, Sweden and New Zealand, have the same participants' behaviour in the simulated hotel fire scenario. This is achieved by comparing the actions performed data between the VR only condition and Silvia et al.'s results.
    \item \textbf{O4}: To evaluate the effect of fire smell on participants' immersion, perception of realism, usage of equipment, and any feeling of distress in VR. For this objective, we compared the questionnaire data between the VR with smell and VR only conditions.
\end{itemize}

\subsection{Experimental Setup}
\begin{figure}[H]
 \centering 
\includegraphics[width=\columnwidth]{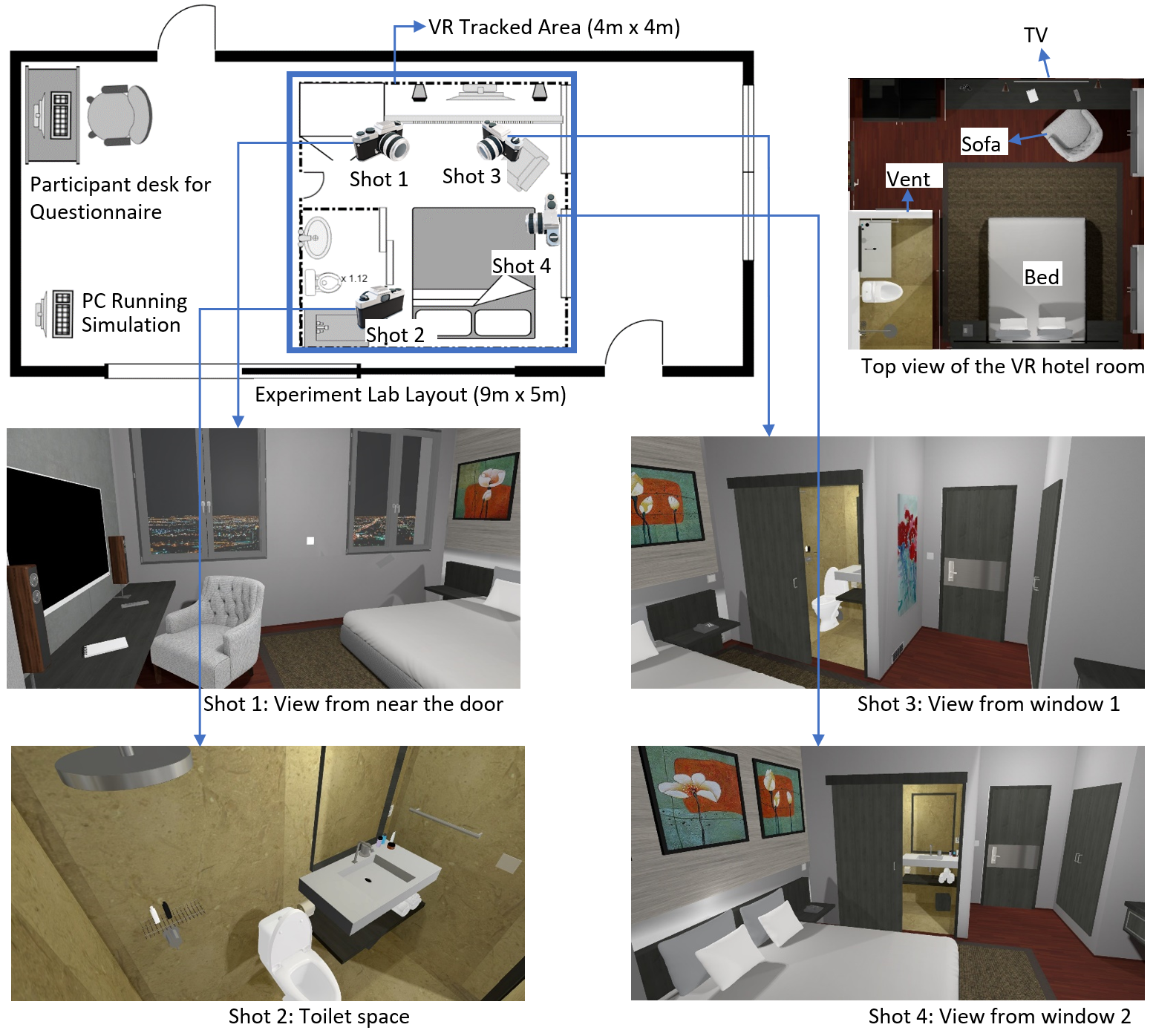}
 \caption{ Lab floor plan identifying the different areas of the experiment space including the VR tracked area. Shot 1-4 are the photos taken in the VR space, and they are also marked on the floor plan for reference.}
 \label{ExptSetup_Layout_VRView}
\end{figure}

The user study was carried out in a lab with a tracking space of 9m by 5m. The experiment room layout along with the tracked area and the experiment setup are illustrated in Figure \ref{ExptSetup_Layout_VRView}. The VR hotel room was kept the same as the original simulation designed by Silvia et al. \cite{arias2019forensic}. Also, the tracking area was kept the same at 4m by 4m. The hotel room consisted of a bedroom with an attached toilet and shower room and had no resemblance with the MGM Grand hotel room in 1980. This was to ensure participants felt the same as they would if they checked in a present-day hotel and in turn exhibited more real-life behaviours that would inform evacuation guidelines.

The hotel room and the attached toilet were furnished with modern-looking accessories. The objects in the hotel room and the VR space can be seen in Figure \ref{ExptSetup_Layout_VRView}. The interaction with the objects was made as close to the reality as possible. Some of the objects were kept static (e.g. bed, desk, bathroom sink, vent, etc.) and the interaction with them was based on collision. For static objects, participants can put other 3D objects on them but they are immovable in the VR space. Some objects were partially movable (e.g. doors (room, toilet, cabinet), windows, closet, toilet seat, faucets (sink, shower), etc.). The interaction with these objects was mainly swing motion pivoted on the hinge or sliding motion to open and close the doors or windows, or in the case of faucets to start and stop the flow of water stream. There were also objects that were completely movable (e.g. pillows, rolled towels, TV remote, phone, waste baskets, speakers, sofa, notepad, etc.). These objects can be picked up with the HTC Vive hand controllers and moved around the space. Some of the movable objects can get wet (pillows, rolled towels, waste baskets) and can be used to block the vent when smoke starts coming out of the vent. There was no smoke in the hotel room at the start of the experiment and participants were able to interact and explore the environment as they pleased. After 6-8 minutes of the exploration phase, the researcher triggered smoke that entered through the vent. 

The objects placed in the hotel room were based on the objects identified by survivors in the NFPA survey \cite{bryan1983review} except for the objects that cannot be realistically simulated in real-time in VR such as bed sheets and curtains. Also, radio was not added to the hotel room as they are generally unavailable in modern hotel rooms. A fluid simulation (water accumulating in wastepaper baskets) was also not included to keep the VR simulation running at a high frame rate.

\subsection{Participants}
There were 40 participants in the experiment. In order to represent the general population, an equal number of male and female participants were recruited - 20 male and 20 female. The age varied between 18 and 47 with an average value of 27.4, standard deviation of 6.9, and mode of 27. To recruit the participants, user study posters were advertised on social media as well as on the university's campus noticeboards. Each participant received a \$10 voucher as an appreciation for their participation. In the questionnaire after the experiment, participants were asked about their training experience in fire safety. Forty percent of the VR with visual-audio-only sample indicated they trained for fire safety. For VR with smell condition, twenty percent of the participants reported having previous training on actions to take in a fire. In the real MGM Grand hotel fire, seventeen percent of the survivors indicated previous training experience.  

\subsection{Equipment}
To render the VR experience, HTC Vive Pro Eye head-mounted display (HMD) was used along with its two hand controllers. The HTC headset features dual OLED 3.5" diagonal with a resolution of 1440 x 1600 pixels per eye (2880 x 1600 pixels combined). It has 110° of field of view, 90 Hz refresh rate, and 6 degrees of freedom (DOF) tracking. To allow more freedom of movement for the participants, the HTC Vive headset was connected wirelessly to the computer running the VR simulation. The HTC headset position and orientation are tracked using inside-out tracking from base stations. We have four base stations 2.0 installed in the lab, covering an area of about 5.0 m x 8.0 m. For our user study, we limited the tracking to 4 m x 4 m space to enable line of sight to most of the base stations which ensured high-quality tracking throughout the experiment. We also had a minimum of 0.5m buffer zone on each side of the wall to prevent participants from bumping into walls, running into furniture, or any other equipment in the lab. In the VR space, virtual boundaries also appeared showing the physical limitation of the operation area. Prior to the main study, participants were exposed to a training environment where they can use the HTC's hand controllers to interact with the virtual object and also get familiar with the physical boundaries of the operation space.  

To create the VR experience, a high-end game computer was used to fulfill the hardware requirements necessary for the HTC headset. The computer contained AMD Ryzen 7 2700X CPU, Nvidia GeForce GTX 2080 8 GB GPU, 32 GB RAM, and was capable of consistently generating nearly 90 FPS for a smooth VR experience. The VR environment was created in Unity 3D game engine version 2017.3.1f1 using existing 3D assets and some made with the SketchUp software. The smell prototype was created using the Prusa i3 MK3S+ 3D printer.

\subsection{Measurements}
Similar to Silvia et al. \cite{arias2019forensic}, participants behaviour was evaluated using five actions. The rationale for choosing five actions was to be consistent with the NFPA survey that also asked the survivors to identify five actions prior to the evacuation \cite{ bryan1983review}. These actions were not disclosed to the participants, and it was at their discretion to perform them during the study. The five actions are as follows:
\begin{itemize}
    \item Turn on TV:  After the fire, did the participant turned on TV to look for news or any other information related to the fire. 
    \item Tap to use phone: Whether the participant picked up phone and tried to contact the hotel service desk to inquire about the smoke or fire.
    \item Block vent: When the smoke started to come out of the vent, did the participant tried to block the vent with objects (such as towel) to stop the flow of smoke inside the room.
    \item Open window: To remove the smoke out of the room, did the participant opened the windows
    \item Signal outside: Did the participant make any attempt to grab attention of rescue services by either waving hands or objects close to the window or calling out loud for help, or throwing objects out of the window.   
\end{itemize}
There were other actions that can be performed during the experiment, such as covering the face with a towel, switching all the water taps on and blocking the sink to let the water out, lying on the floor, and turning off the lights. These actions can be performed but they are out of the scope of this paper. They were not considered to assess behavioural realism.

Apart from the behavioural realism, we also evaluated the perception of participants on immersion, the realism of the virtual scenario and the environment including smell, and if they experience any physical discomfort, stress, fear, disorientation, or insecurity. Table \ref{Table_Questionnaire-Questions} shows the questions asked. The questions were based on the prior work \cite {arias2019forensic} and were mixture of 7-point Likert scale ratings and textual input. The Likert scale ratings were presented as a radio-button grid format. 

\begin{table}[H]
 \centering 
 \caption{Post-session questions in the qualtrics questionnaire to measure immersion, realism, physical discomfort, stress, fear, disorientation, or insecurity.}
 \includegraphics[width=\columnwidth]{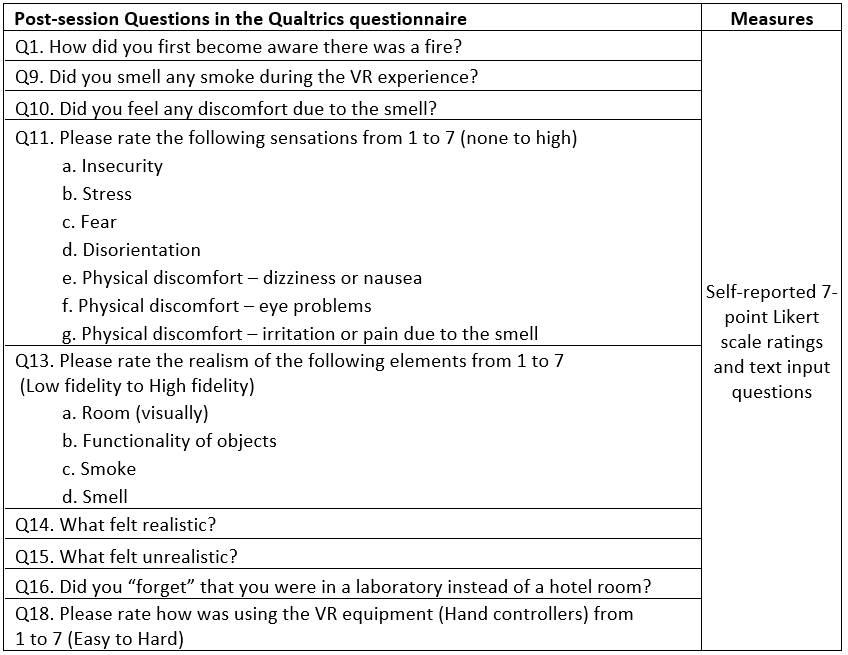}
 \label{Table_Questionnaire-Questions}
\end{table}

\subsection{Pilot Testing}
Before doing the main experiment, we carried out pilot testing on five volunteers to identify any issues with the system, experiment procedure, or questionnaire. During pilot testing, participants experienced the same condition as designed for the main experiment and answered the questionnaire afterwards. We did the screen recording of the virtual environment experienced by the volunteers for behavioural analysis afterwards. The behaviours were coded afterwards using the recorded video. There were few issues highlighted during the pilot testing, such as some participants finding it hard to interact with the door and windows in the VR environment. To make the interaction more smoother, windows and doors colliders were resized which allowed easier collision detection in turn improving interaction. An issue with the smoke density shader was also identified during the testing. The shader was not being initialised correctly that lead to the smoke layer covering everything above certain prior to the start of the emergency scenario. This issue was fixed as well. Overall the pilot study improved our system, procedure and questionnaire. 

\subsection{Procedure}
During the recruitment, participants were informed that they would experience a highly realistic hotel room and be asked about their experience. No mention of the fire emergency scenario was made to prevent influencing participants' behaviour and decision-making. Participants secured their time slots through an online booking website that described potential risks and what is involved in the experiment without revealing the true intent. On arrival, each participant received an information sheet and consent form to review and sign as mandated by the ethics committee. After they consented, they were introduced to the VR equipment and informed about the training session (3 to 5 minutes long) prior to the main experiment to get familiar with VR and the interaction with virtual objects. The physical limitation of the space with the virtual lines appearing was also shown during training. On finishing the training session, participants were guided to the starting position of the experiment. Then the VR hotel simulation was started and participants were given a short brief that they have checked into a hotel room to stay for a night. They were also informed that the hotel is highly realistic, and they can interact with virtual objects as they would in real-life. They were told that at the end of the experiment, they would be answering a questionnaire on how realistic it felt. They were also informed that they can explore the room as long as they would like or if they were instructed to stop. Participants were advised that if they experienced any nausea or sickness they should inform the researcher, and they can stop the experiment at any time for any reason. After the brief, the researcher instructed the participant to start the experiment.

\begin{figure}[htp]
 \centering 
\includegraphics[width=0.9\columnwidth]{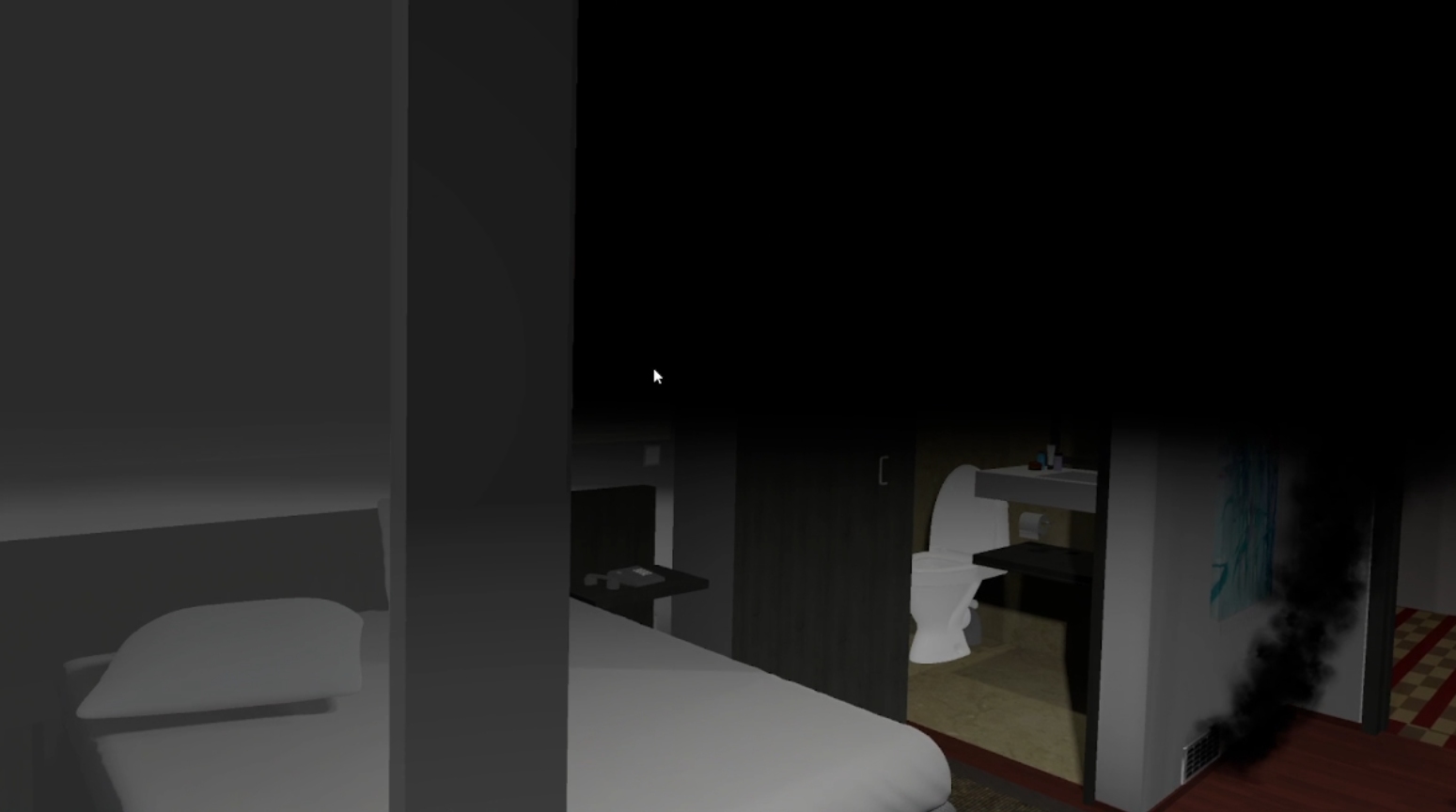}
 \caption{Smoke coming out of the vent and getting accumulated in the hotel room that highly reduced visibility.}
 \label{SmokeAccumulation_DecreaseVisibility}
\end{figure}

Following the instructions, the participants started the experiment. Their view in the virtual reality including the actions was recorded as a video using screen capturing software. During the pilot study, it was observed approximately 6 to 8 minutes proved adequate for participants to explore the VR environment and be able to interact with most of the objects in the hotel room. This time frame established from the pilot study was adopted for the main experiment. When the exploration phase ended (time threshold reached) and the participants were not directly looking at the vent, the smoke was triggered. The smoke plume came out of the vent and ascended towards the ceiling. For the smell condition, the arm of passive olfactory display (POD) came down and the lid covering the fire smell opened up. Participants registered the presence of smoke at varying intervals subsequent to the initial smoke trigger. The time of detection varied depending on their spatial location and orientation in the room at the time of trigger and if there was smell present or not. Most of the participants performed actions to stop or reduce the smoke. In the instance where they were not able to stop or reduce the smoke, it started accumulating in the room causing a gradual reduction in visibility as the smoke layer thickened ceiling down until the visibility was quite low. Fig. \ref{SmokeAccumulation_DecreaseVisibility} illustrates this phenomenon. Importantly, there was no suggestion or any input from the researcher on what to do or what actions to take.

After the interaction with the environment, the experiment was stopped based on participant's actions that fall into three stopping conditions. The first  was that the participants did not do anything to stop the smoke and the visibility decreased with smoke layer reaching the floor level. It was hard to move at this point. Participants were given some time once the smoke reached the floor level to see if they would do anything about the smoke, but most of them stopped moving at this point. The second stopping condition was the participants who were able to stop or reduce the smoke did not do anything else for some time and were anticipating the end of the experiment with a rescue. The third stopping condition was the participants performing the same actions repeatedly. The experiment was terminated based on these conditions as there was no more meaningful data to be gathered. This was consistent with the behaviour of the survivors, who at some point during their several hours in the room ran out of actions to perform and waited for rescue. The data recorded during the experiment included the experiment start time, smoke triggered time, the time when the participant was first aware of the smoke, and the experiment stop time. 

On completion, participants were assisted to remove the VR headset and guided to a computer-equipped desk for a Qualtrics-based questionnaire. A short interview followed to discuss any observed unusual behaviour during the experiment. Instances like a participant locking herself in the toilet or manipulating water taps were probed for rationale. A debriefing session at the end clarified the study's purpose, data usage, and provided compensation vouchers. Participants were then guided to the building exit.

\section{Results} 
This section presents the results of the VR experiment. It starts with the evaluation of the realistic behaviour (the five actions) exhibited by the participants during the experiment. The data from the five action items are compared to the survivors’ behaviour data recorded in the NFPA survey \cite{bryan1983review} as well as to Silvia et. al.’s data \cite{arias2019forensic}. In the second part, the VR-related results show participants’ perception of the VR environment and smell in terms of realism. It also reports if the experiment induced any stress, fear, insecurity, or discomfort. Lastly, the qualitative observations are presented.

\subsection{Human Behaviour}
\begin{figure}[htp]
 \centering 
\includegraphics[width=\columnwidth]{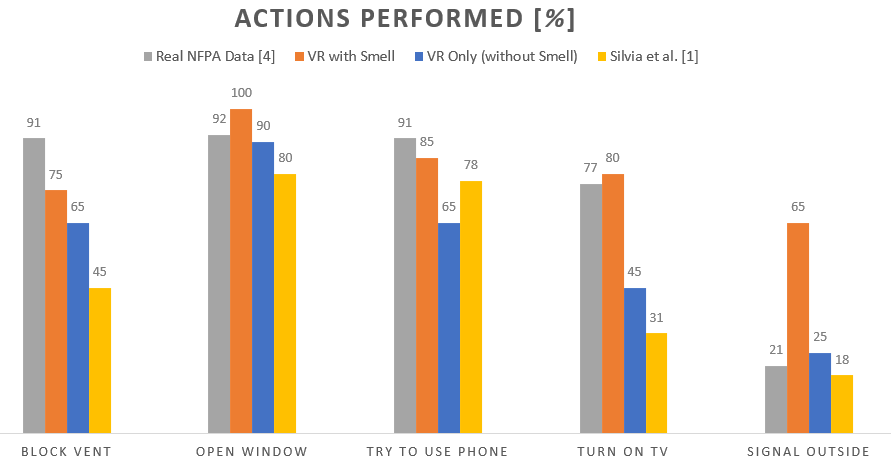}
 \caption{Percentage of the 40 participants performing the five actions with VR only, VR with Smell as well as comparison to Silvia et al.'s and real NFPA data. }
 \label{Results_ActionsPerformed}
\end{figure}

We evaluated participants’ behaviour during the hotel fire based on the five survivor action items identified in the NFPA survey. To examine our experimental results (VR with smell and VR only conditions) with real survivors' NFPA data and Silvia et al.'s results, we computed the percentage of each action and plotted the values to give an overall comparison. Figure \ref{Results_ActionsPerformed} shows the plotted values of our results (VR with smell and VR only conditions), real survivors' NFPA, and Silvia et al.'s data for the five actions. The results in Figure \ref{Results_ActionsPerformed} indicate that the VR with smell condition matched more closely to the real survivors' NFPA data for the five actions except for the “Signal Outside” case where a higher percentage of people in VR performed the action than the real NFPA case. The results of VR only condition for actions performed were more similar to Silvia et al.’s data in spite of the different population demographics (our sample was from New Zealand and Silvia’s from Sweden). The actions performed by the participants were entirely at their discretion and no participants were told or hinted in any way as to what actions they should carry out during the experiment.

To explore the statistical significance and assess whether the results of the two experimental conditions (VR with smell and VR only) address our research objectives, we performed Fisher’s exact tests with Bonferroni correction. For \textbf{O1}, we compared each condition individually (the observed data) to the real survivors' NFPA data (the expected data). Table \ref{FisherExactTest_VR_with_Smell} shows the results of Fisher’s exact test comparing VR with smell condition to the NFPA data. We found only “signal outside” action to be significantly different than the real NFPA results. The other four actions were found to be not significantly different from the real survivors' NFPA data. Table \ref{FisherExactTest_VR_only_Without_Smell} compares the VR Only condition to the NFPA data using Fisher’s exact test. From Table \ref{FisherExactTest_VR_only_Without_Smell}, we can see “block vent”, “try to use phone”, and “turn on tv” actions were significantly different than the real survivors' NFPA data. For \textbf{O2}, Fisher’s exact test was performed between the two experimental conditions (VR with smell and VR Only) to find the change in behaviour due to the smell. The results are in Table \ref{FisherExactTest_VR_with_Smell_n_VROnly}. We found only one action “signal outside” to be significantly different. For \textbf{O3}, Fisher’s exact test was performed between VR only condition and Silvia et al.'s data to understand if there was a change in behaviour between the two populations. The results are in Table \ref{FisherExactTest_VROnly_Silvia_NZ_SwedishData}. We did not find any action performed to be significantly different between New Zealand and Swedish populations.

\begin{table}[H]
 \centering 
 \caption{Fisher’s exact test results comparing VR with smell condition with real survivors NFPA data (significance level is  $< 0.0125$ with Bonferroni correction applied $(0.05/4)$)}
 \includegraphics[width=\columnwidth]{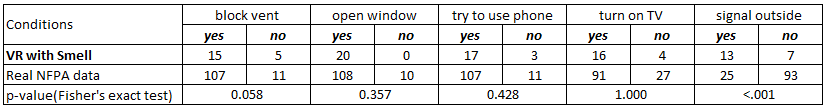}
 \label{FisherExactTest_VR_with_Smell}
\end{table}

\vspace{-3em}

\begin{table}[H]
 \centering 
 \caption{Fisher’s exact test results comparing VR only (without smell) condition with real survivors NFPA data (significance level is $< 0.0125$)}
 \includegraphics[width=\columnwidth]{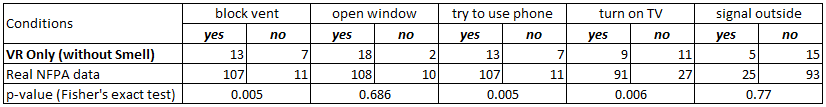}
 \label{FisherExactTest_VR_only_Without_Smell}
\end{table}

\vspace{-3em}

\begin{table}[H]
 \centering 
 \caption{Fisher’s exact test results comparing VR with smell condition with VR Only condition (significance level is $< 0.0125$)}
 \includegraphics[width=\columnwidth]{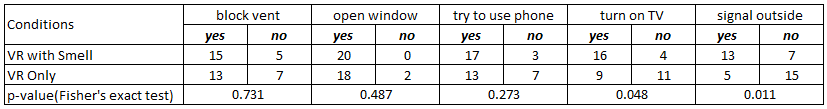}
 \label{FisherExactTest_VR_with_Smell_n_VROnly}
\end{table}

\vspace{-3em}

\begin{table}[H]
 \centering 
 \caption{Fisher’s exact test results comparing VR Only condition (New Zealand population) with Silvia et al.'s data (Swedish population) (significance level is  $< 0.0125$)}
 \includegraphics[width=\columnwidth]{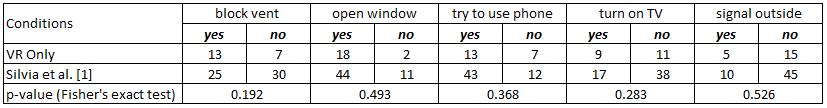}
 \label{FisherExactTest_VROnly_Silvia_NZ_SwedishData}
\end{table}
 
\vspace{-1em}

\subsection{Perception of the VR Experience} \label{section_VR_Perception}
Participants' perception of the VR experience was recorded using a questionnaire after the experiment. The results of the participant's responses are summarised in this section.

\subsubsection{Realism} \label{section_Realism}
\begin{figure}[H]
 \centering 
\includegraphics[width=0.9\columnwidth]{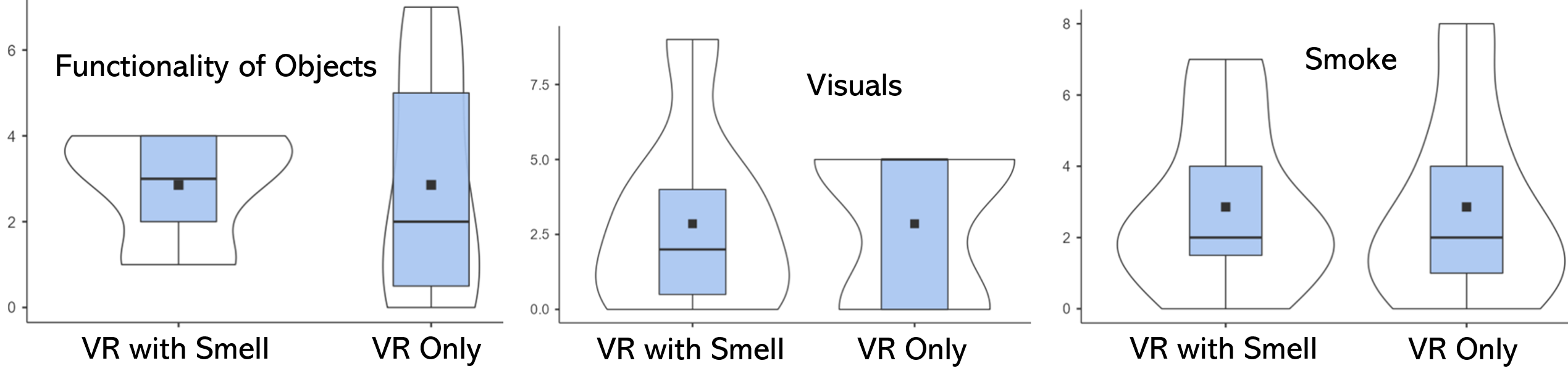}
 \caption{Violin plots of participants rating from 1 (low) to 7 (high) on the realism of different components of the VR experience. These include: the visual fidelity of the hotel room, objects in the space functionality, and smoke's visual fidelity.}
 \label{VR_relatedResults_Realism}
\end{figure}

 In the questionnaire, participants were asked to rank the realism of the VR environment,the functionality of the objects, smoke, and smell from 1 (low) to 7 (high). The range of their responses for VR with smell condition as well as for VR only (without smell) condition are shown in Figure \ref{VR_relatedResults_Realism} except for the smell realism. The realism of smell being closer to the fire smell can only be rated for the VR with smell condition. Among the realism ratings, for VR with smell condition, the smell was rated the highest with the mean value of 5.9 and standard deviation of 1.29, then the visual fidelity of the VR environment(the hotel room) with the mean value of 5.35 and standard deviation of 1.09. Realism of the smoke was rated with the mean value of 5.2 and standard deviation of 1.40, and the functionality of the objects with the mean value of 4.65 and standard deviation of 1.79. For the VR only condition, the realism for VR environment (the hotel room), functionality of objects, and smoke appearance were rated with the mean values of 5.50, 5.15, 5.35, and standard deviation of 1.15, 1.14, and 1.31, respectively. To find if there was a significant difference between the VR with smell and VR only (without smell) conditions, we applied the Mann–Whitney U test and found no significant difference. The resulting p-values for VR environment (the hotel room), functionality of objects, and smoke appearance were $p = 0.74$, $p < 0.41$, and $p < 0.76$.

\subsubsection{Immersion}
\begin{figure}[H]
 \centering 
\includegraphics[width=0.4\columnwidth]{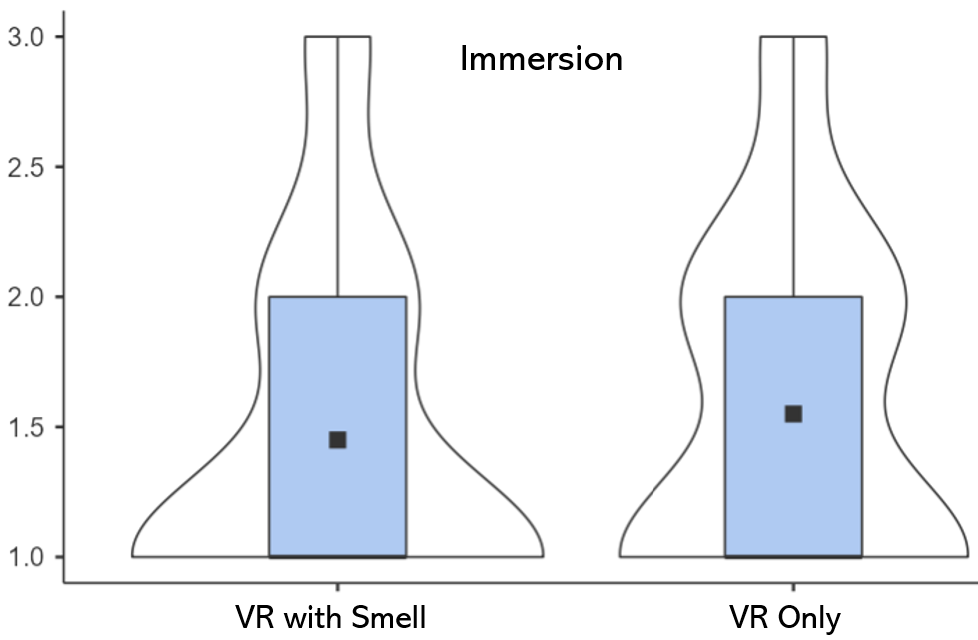}
 \caption{Violin plots of participants in response to the question “Did you “forget” that you were in a laboratory instead of a hotel room?”. Participants in the “sometimes” category gave their own description of the immersion level. }
 \label{VR_relatedResults_Immersion}
\end{figure}

For immersion, we asked the participants "Did you “forget” that you were in a laboratory instead of a hotel room?". There were given three options, "yes", "no", or add their own answer. Figure \ref{VR_relatedResults_Immersion} shows the results for both conditions. The participants who chose their own answer sometimes felt completely immersed while at other times they were aware of the surroundings. For the immersion condition, VR with smell condition had 13 participants feeling immersed and VR only condition had 11 participants feeling immersed. To find if there was a significant difference between the VR with smell and VR only (without smell) conditions, we applied Fisher's exact test and found, $p = 0.89$, no significant difference.

\subsubsection{Use of Hand Controller}
\begin{figure}[H]
 \centering 
\includegraphics[width=0.4\columnwidth]{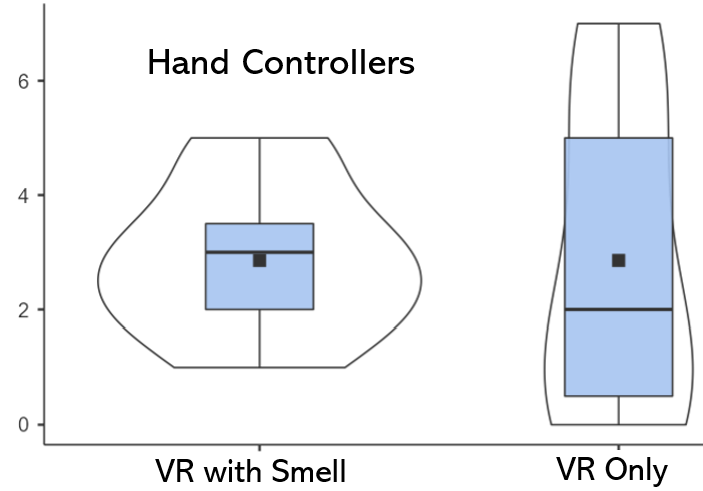}
 \caption{Violin plots of participants rating how easy it was to use the hand controllers from 1 (easy) to 7 (hard). }
 \label{VR_relatedResults_HandControllers}
\end{figure}

For hand controllers, we asked the participants to indicate the level of difficulty to use the controllers for interaction from 1 (easy) to 7 (hard). The summary of their responses for both conditions are presented in Figure \ref{VR_relatedResults_HandControllers}. For VR with smell condition, the mean value was 3.5 and standard deviation 0.34, and for VR only condition, the mean value was 3.2 and standard deviation 0.4. This indicated hand controller were relatively easier to use for interaction in the virtual environment. To find if there was a significant difference between the VR with smell and VR only (without smell) conditions, we applied the Mann–Whitney U test and found, $p = 0.78$, no significant difference.

\subsubsection{Discomfort and Feelings}
\begin{figure}[H]
 \centering 
\includegraphics[width=0.8\columnwidth]{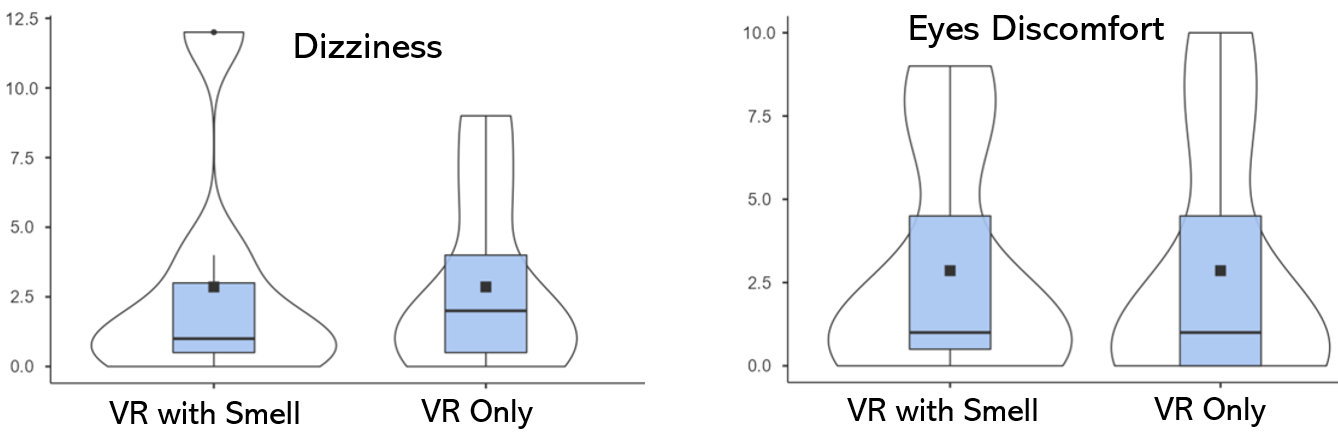}
 \caption{Violin plots of participants rating the discomfort experienced during the VR experiment, from 1 (low) to 7 (high).}
 \label{VR_relatedResults_Discomfort}
\end{figure}

Additionally, participants were asked to rate their level of discomfort from 1 (low) to 7 (high). The rating was for any experience of dizziness or eye discomfort. The results of their responses are shown in Figure \ref{VR_relatedResults_Discomfort}. To find if there was a significant difference between the VR with smell and VR only (without smell) conditions, we applied the Mann–Whitney U test and found, $p = 0.48$ for dizziness and $p = 0.29$ for eyes discomfort, no significant difference.

\begin{figure}[H]
 \centering 
\includegraphics[width=\columnwidth]{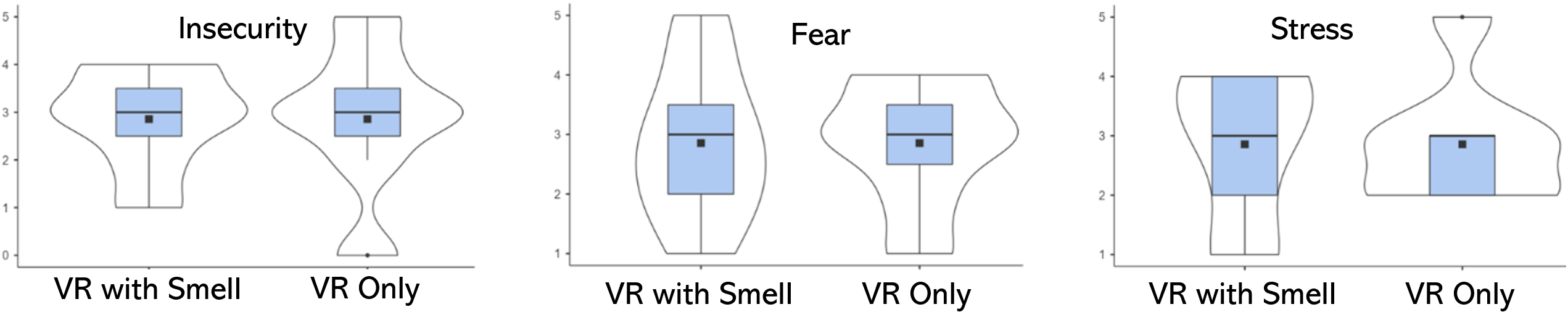}
 \caption{Violin plots of participants rating their feelings of fear, stress, and insecurity experienced during the experiment due to the emergency, and not because they are in an experiment.}
 \label{VR_relatedResults_ReportedFeelings}
\end{figure}

Participants were also asked to rate their feelings of insecurity, stress, and fear from 1 (none) to 7 (high). Figure \ref{VR_relatedResults_ReportedFeelings} summarises the results for both between-subject conditions. To find if there was a significant difference between the VR with smell and VR only (without smell) conditions, we applied the Mann–Whitney U test and found no significant difference ($p = 0.46$ for insecurity, $p = 0.90$ for stress, and $p = 0.50$ for fear).

\subsection{Qualitative observations in VR}
We also observed participants' behaviours during the experiment, and if any unusual action was noticed, participants were asked about the rationale of their actions in an informal interview after the experiment. Once the smoke started, the sound of firetrucks and ambulances also started coming from the window. A lot of participants moved toward the window to understand what is happening outside. However, when they saw the smoke coming out of the vent, there was a sense of urgency. A lot of participants went towards the door and opened the door to see if they can escape. There was a lot of smoke visible in the hallway, and they were reminded by the researcher at this point that the "outside is worse". On hearing the researchers, they returned to the room and on the way back they saw the windows. A lot of the participants moved to the windows to open them to let the smoke out. Some people opened them and closed them immediately. On asking about their reasoning for the action, they argued that they were scared that the action would lead to more oxygen inside the room which could possibly result in an explosion or more fire. After opening the windows, some participants started to control the smoke coming from the vent. Once they saw that the smoke layer started to recede towards the ceiling, the feeling of urgency decreased. At this point, they also started looking for news on the TV, tried calling the reception or police on the phone, and some people started waving near the window or throwing objects from the window.  

Some of the less common behaviours observed during the study were as follows. Some participants looked for a fire extinguisher. Some made the towel wet and used that to cover their face and not put it on the vent. Three participants got on the floor and waited for emergency staff to come to rescue them; one of them also rolled on the floor. The reasoning given by them was their prior training during school which taught them to stop, drop, and roll or take cover in the case of a fire. Two participants locked themselves in the toilet and one of them also switched on the shower. A few participants tried to fill the paper baskets with water to extinguish the fire. This was reasonable action; however, fluid motion was not included in the VR simulation to avoid dropping the frame rates below 90 FPS. 

A small number of participants were not able to control the smoke. They clearly saw the smoke but did not try to control it. Some of the participants tried to disperse it with a pillow or other objects; however, they did not perform actions (blocking the vent, or opening windows) to reduce the smoke or stop it at the vent. Once the smoke accumulated in the room, these participants stopped moving as they were not able to see anything. At this point, the researcher stopped the experiment.

\section{Discussion}
We evaluated the \textbf{O1} using Fisher's exact tests results in Table \ref{FisherExactTest_VR_with_Smell} and Table \ref{FisherExactTest_VR_only_Without_Smell}. The results for VR with smell condition (Table \ref{FisherExactTest_VR_with_Smell}) showed the participants performing four (“block vent”, “open window”, “tap to use phone”, and “turn on tv”) out of five actions with no significant difference than the real survivors NFPA results. In comparison, participants with VR only condition (Table \ref{FisherExactTest_VR_only_Without_Smell}) performed two (“open window” and “signal outside”) out of five actions similar to the real survivors' NFPA data. This indicates that the introduction of fire smell in a hotel fire evacuation scenario increases the realistic behaviour exhibited by the participants. The observed difference in the number of actions closer to the real fire scenario indicates that olfactory cue could impact human decision-making during fire evacuation scenarios. Smell stimuli can trigger a cognitive response associated with danger and urgency, leading to a more appropriate response taken by the evacuee. The “signal outside” action for VR with smell condition was even at a higher percentage than the real survivors' data (Figure \ref{Results_ActionsPerformed}). A possible explanation for the significant difference (Table \ref{FisherExactTest_VR_with_Smell}) and a higher percentage in results for the “signal outside” action is the presence of a smell-only cue in the environment. That could have resulted in participants perceiving the situation as more dangerous. For the real scenario, apart from the smell, there were also haptic cues that could have resulted in different threat perception. These findings align with the predictive behavioural model presented by Kuligowski which suggests that certain cues (physical and social) increase the chances of an individual taking actions during a building fire \cite{kuligowski2009process}. 

To assess the \textbf{O2}, Fisher's exact test was performed between VR with smell and VR only condition. The results (Table \ref{FisherExactTest_VR_with_Smell_n_VROnly}) showed a significant difference for only one action "signal outside". This indicates that the introduction of fire smell could possibly change participants' behaviour. The action that was different involved seeking information or help from outside. For \textbf{O3}, the VR Only condition results were compared with Silvia et al.'s data using Fisher's exact test. As can be seen in Table \ref{FisherExactTest_VROnly_Silvia_NZ_SwedishData} there was no significant difference in actions performed between the two populations. This suggests that there is a similar emergency response between the two demographics during a hotel fire scenario. The \textbf{O4} was analysed using the findings in section \ref{section_VR_Perception}. The results indicated no significant difference between VR with smell and VR only conditions for the realism of the VR environment, realism of smoke and functionality of objects, immersion, use of hand controllers, discomfort, and reported feelings (section \ref{section_Realism}).

The hotel room was designed to resemble a modern hotel room and was furnished with all the items that would allow the participants to perform similar actions as the survivors. The room had a toilet with a sink and a shower with running water. It also contained a TV with remote, soakable pillows and towels, paper baskets, a sofa, cabinets, light switches, speakers and interactable doors and windows. For this study, we only looked at five actions that were also shown in the NFPA survey \cite{bryan1982human}. There were also other actions carried by the participants. These actions include participants trying to remove the vent grill and throwing towel to control the smoke, or switching off the lights with the intent to stop an electrical short circuit and secondary fire. Some actions taken by the participants were categorised under the five studied actions. For example, "to signal outside" some participants switched lights on and off multiple times while others waved objects near the window, and some also called out loud for help. All of these actions were classified as "to signal outside". Overall limited actions have been studied in this experiment to ensure conditions are controllable. The study can be expanded to more actions in the future. Also, the order and sequence of actions performed by the participants can be investigated. 

Another important factor that might have influenced participants' actions during the experiment is their present-day understanding or expectations of what should be there in a hotel room such as a smoke alarm, floor layout with evacuation routes behind the door, audible loudspeaker fire announcement, visible fire exit signage, and sprinkler system. Familiarity with these safety systems would lead to different actions than the ones performed by the survivors in 1980. The population demographics of our experiment were also different from the survivors. The mean age of the participants for VR with smell condition was 27, and for VR only condition was 28 whereas for the real NFPA data, the mean age was 46. A lack of sense of anxiety was observed in the participant during the experiment which might not be the case in an actual fire scenario. There was a sense of safety and lack of consequences among the participants when they were doing the experiment. These differences could also lead to different behaviours in the virtual world.


\vspace{2mm}
\noindent \textbf{Limitations and Future Directions}

\noindent Smell dispersion in air is a complex dynamic phenomena, and it is hard to simulate it to a similar fidelity as the real-world. Our smell delivery system, POD, does not simulate the spread of smell in a three dimensional environment as the real-world. However, from the results (Figure \ref{VR_relatedResults_Realism}) we can see that the perception of smell by the participants is closer to what they expected, and it is aligned with the smoke visuals. As our focus is human behaviour, the smell dispersion does not have to match the real-world. The smell should allude to fire or smoke, and should not be completely different to the visual impression as that could lead to break in presence. Some of the participants reported that they expected a change in the smell intensity when they went far from the smoke source. Change in smell intensity and its effects on immersion can be studied in the future. Apart from smell, social cues and the inclusion of other sensory stimuli (heat, wind) are possible future research directions.

 Social cues played an important role during the evacuation behaviour as identified in the real NFPA survey \cite{bryan1983review} and in Kinateder's work \cite{kinateder2012social}. Social influence was not studied in this experiment to avoid adding another control variable and should be separately investigated. Social influence can be added in the future by either introducing a non-player character (NPC) and studying the evacuation behaviour, or adding in one or two people in the room using avatars. 
 
 The heat sensory cue should be investigated in the future. Similar to the olfactory cue, heat is an important sensory stimuli during fire \cite{arias2019forensic, kuligowski2008modeling}, and it can be added in conjunction with smoke accumulation. The effect of heat on participants' realistic behaviour should be quantified. While the fidelity of the heat simulation does not have to match the real-world heat dissipation, it should allude to the sensory stimuli associated with heat. Additionally, to gain further insights into the reasons behind performing certain actions in the VR environment, addition measures, such as heart rate variability and galvanic skin conductance should  be added. These measures will complement the recorded behaviour (actions performed) data. The limitations identified in this section provide potential directions for future research.

\section{Conclusion}
Virtual Reality (VR) allows the simulation of dangerous situations, such as hotel fires, with a high level of fidelity while ensuring safety and practicality. Previous research using VR in fire evacuation scenarios indicated that people performed similar actions in VR as the real-life scenarios. However, some of the actions differed in the VR case from the real-life. Researchers argued that the difference in behaviour (actions performed) might be the result of missing sensory cues in VR. Our work introduced a fire-related olfactory cue in the VR simulation of a hotel fire and examined how it affected participants' actions. We conducted a between-subject user study with two conditions: one with VR displaying the smell stimuli during the fire, and the other being VR only condition (without the smell stimuli). We observed participants’ behaviour without alluding to them any potential actions to take. Our results indicated that the presence of an olfactory cue as per the event happening in virtual reality increases the behavioural realism shown by the participants. However, the perception of the event and virtual environment was similar between the two conditions.  

\vspace{4mm}
\noindent \textbf{Ethical Consideration}
\vspace{1mm}

\noindent Prior to conducting the experiment, the study and procedures were reviewed and approved by the UC Human Research Ethics Committee (HREC Ref: 2022/38/LR-PS). All participants provided their informed consent before starting the user study. 


\acknowledgments{
The authors would like to acknowledge Silvia Arias and Jonathan Wahlqvist from Division of Fire Safety Engineering at the Lund University for providing the MGM Grand Hotel fire simulation. Special thanks to Conan Fee and Adrian Clark from the School of Product Design at the University of Canterbury for valuable guidance during the development of the olfactory display. The authors would also like to acknowledge Rory Clifford from Human Interface Technology Laboratory at the University of Canterbury for acquiring and testing the essential oils to ensure an appropriate smell for the user study.}

\bibliographystyle{unsrt}

\bibliographystyle{abbrv-doi}

\bibliography{references}
\end{document}